\documentclass[a4paper,12pt]{article}

 \usepackage{amsfonts}
\usepackage{verbatim}

\newcommand\ee{\end{eqnarray}}      
\newcommand\be{\begin{eqnarray}}

\begin{document}

\begin{flushright}
SISSA/49/2007/EP\\ hep-th/0707.3922 
\end{flushright}

\vskip 0.5cm

\begin{center}

{\Large \bf BRST, anti-BRST and gerbes}\\

\vskip 1cm

{\bf L. Bonora, R. P. Malik\footnote {On leave of absence from
Centre of Advanced Studies, Physics Department, Faculty of
Science, Banaras Hindu University, Varanasi-221 005 (U.P.),
India.}}\\ {\it International School for Advanced Studies
(SISSA),}\\ {\it Strada Costiera, Via Beirut n.2-4, 34013 Trieste,
Italy}\\

and\\

{\it INFN, Sezione di Trieste, Trieste, Italy}\\ {\small {\bf
E-mails: bonora@he.sissa.it; malik@bhu.ac.in}}\\

\end{center}

\vskip 4cm

\noindent {\bf Abstract:} We discuss BRST and anti--BRST
transformations for an Abelian antisymmetric gauge field in 4D and
find that, in order for them to anticommute, we have to impose a
condition on the auxiliary fields. This condition is similar to
the Curci--Ferrari condition for the 4D non--Abelian 1-form gauge
theories and represents a consistency requirement. We
interpret it as a signal that our Abelian
2-form gauge field theory is based on gerbes. To support
this interpretation we discuss, in particular, the case of the
1--gerbe for our present field theory and write the relevant
equations and symmetry transformations for 2--gerbes.

\vskip 2cm



\newpage

\section{Introduction}

While the BRST symmetry has been a fundamental tool for the study
of quantum field theories in the last three decades, the
anti--BRST symmetry does not seem to have played more than a
decorative role. In this paper, we would like to suggest that,
perhaps, the system of BRST and anti--BRST symmetry contains more
information than it is generally attributed to it. To start with,
we discuss some features of BRST and anti--BRST transformations in
theories of free Abelian two--form $B_{\mu\nu}$ fields and show that the
requirement that the BRST and anti-BRST transformations must
anti-commute, imposes a constraint on the fields of the theory,
very similar to the one that must be imposed for the 4D 1-form
non-Abelian gauge theories (i.e. the Curci-Ferrari condition). The
idea we want to develop in this paper is that this type of
constraints is characteristic, not only of the non-Abelian gauge
theories, but also of higher form Abelian gauge theories
whose field content is based on gerbes. To clarify this point, we
first give a brief introduction to the subject of the Abelian
gerbes. Then, we find the BRST and anti-BRST transformations for
the 1-gerbe and show that their anti-commutativity requires
precisely constraints of the above type. We show also that the
2-gerbes require two such constraints. We interpret all this as evidence
that, indeed, such type of Curci-Ferrari constraints are
characteristic of the gerbe-based field theory.

The contents of the paper are organized as follows. In section 2,
we discuss the bare essentials of the nilpotent but
non-anti-commuting symmetry transformations for the free Abelian
2-form gauge theory in 4D. This is followed, in section 3, by the
discussion and derivation of the nilpotent and anti-commuting
(anti-)BRST symmetry transformations for the above theory. Section
4 deals with the brief synopsis of the Abelian gerbes. The
material of section 5 concerns the anti-commutativity property of
the (anti-)BRST symmetry transformations and its connection with
the gerbes. Finally, we make some concluding remarks in section 6.

\section{Nilpotent and non-anticommuting symmetry transformations:
a brief synopsis}

\noindent We begin with the following nilpotent symmetry invariant
Lagrangian density for the 4D free Abelian 2-form gauge theory
[1,2,3]
\begin{eqnarray}
{\cal L}_b &=& \frac{1}{12} H^{\mu\nu\kappa} H_{\mu\nu\kappa} +
B^\mu (\partial^\nu B_{\nu\mu} - \partial_\mu \phi) -
\frac{1}{2} B^\mu B_\mu - \partial_\mu \bar\beta \partial^\mu \beta\nonumber\\
&+& (\partial_\mu \bar C_\nu - \partial_\nu \bar C_\mu) (\partial^\mu C^\nu)
+ \rho\; (\partial \cdot C + \lambda) + (\partial \cdot \bar C + \rho)\;
\lambda.
\end{eqnarray}
In the above, the kinetic energy term is constructed with the
curvature tensor $H_{\mu\nu\kappa}$ which is an intrinsic
component of the three--form $H^{(3)} = d B^{(2)} = (1/3!) (dx^\mu
\wedge dx^\nu \wedge dx^\kappa) H_{\mu\nu\kappa}$ where 2-form
$B^{(2)} = (1/2!) (dx^\mu \wedge dx^\nu) B_{\mu\nu}$ defines the
gauge potential $B_{\mu\nu}$ of the theory \footnote{We follow
here the convention and notations for the 4D Minkowski spacetime
manifold with the metric $\eta_{\mu\nu} =$ diag $(+ 1, - 1, -1,
-1)$ where the Greek indices $\mu, \nu....= 0, 1, 2, 3$.}. The
Nakanishi-Lautrup auxiliary vector field $B_\mu$ is invoked to
linearize the gauge-fixing term $[(1/2) (\partial^\nu B_{\nu\mu} -
\partial_\mu \phi)^2]$. The latter requires, for the nilpotent
symmetry invariance in the theory, the fermionic vector
(anti-)ghost fields $(\bar C_\mu)C_\mu$ as well as the bosonic
(ghost-for-ghost) fields $(\bar\beta)\beta$. The above symmetry
invariant Lagrangian density also requires fermionic auxiliary
ghost fields $\rho$ and $\lambda$ (for the gauge-fixing of the
vector (anti-)ghost fields) and a massless (i.e. $\Box \phi = 0$)
scalar field $\phi (x)$ for the stage-one reducibility that is
present in the Abelian 2-form gauge theory.

The following off-shell nilpotent, local, covariant, continuous
and infinitesimal transformations\footnote{It will be noted that
these nilpotent transformations are same as the ones given in [4].
These differ from our earlier choice of the same [2,3] by a sign
factor.}
\begin{eqnarray}
&& \tilde s_b B_{\mu\nu} = - (\partial_\mu C_\nu - \partial_\nu C_\mu),
\qquad \tilde s_b C_\mu = - \partial_\mu \beta, \qquad \tilde s_b
\bar C_\mu = - B_\mu, \nonumber\\
&& \tilde s_b \phi = \lambda, \qquad \tilde s_b \bar \beta = - \rho,
\qquad \tilde s_b \bigl [\rho, \lambda, \beta, B_\mu, H_{\mu\nu\kappa} \bigr ]
= 0,
\end{eqnarray}
\begin{eqnarray}
&& \tilde s_{ab} B_{\mu\nu} = - (\partial_\mu \bar C_\nu -
\partial_\nu \bar C_\mu), \qquad
\tilde s_{ab} \bar C_\mu = + \partial_\mu \bar \beta,
\qquad \tilde s_{ab}  C_\mu = + B_\mu, \nonumber\\
&& \tilde s_{ab} \phi = \rho, \qquad \tilde s_{ab} \beta =
- \lambda, \qquad \tilde s_{ab}
\bigl [\rho, \lambda, \bar\beta, B_\mu, H_{\mu\nu\kappa} \bigr ] = 0,
\end{eqnarray}
leave the above Lagrangian density (1) quasi-invariant because it
transforms as: $\tilde s_b {\cal L}_b = - \partial_\mu [ B^\mu
\lambda + (\partial^\mu C^\nu - \partial^\nu C^\mu) B_\nu - \rho
\partial^\mu \beta]$ and $\tilde s_{ab} {\cal L}_b = -
\partial_\mu [B^\mu \rho + (\partial^\mu \bar C^\nu -
\partial^\nu \bar C^\mu) B_\nu - \lambda \partial^\mu \bar
\beta]$. These transformations have been christened as the
(anti-)BRST symmetry transformations $\tilde s_{(a)b}$ for the 4D
free Abelian 2-form gauge theory [2,3]. However, there is one key
property that is {\it not} satisfied by the above nilpotent
symmetry transformations: the BRST ($\tilde s_b$) and anti-BRST
($\tilde s_{ab}$) transformations do not anti-commute (as is the
case, for instance, with the true (anti-)BRST symmetry
transformations found in the case of any arbitrary gauge (or
reparametrization) invariant theories that are endowed with the
first-class constraints in the language of the Dirac's
prescription for the classification scheme [5,6]).

In fact, while this is true for most of the local fields, present
in the Lagrangian density (1), namely;
\begin{eqnarray}
\{ \tilde s_b, \tilde s_{ab} \} \; \Phi (x) = 0, \quad \Phi (x) =
B_{\mu\nu}, B_\mu, \beta, \bar \beta, \lambda, \rho,
\end{eqnarray}
it can be readily checked, from (2) and (3), that $\tilde s_{ab}
\tilde s_b \bar C_\mu = 0, \tilde s_b \tilde s_{ab} C_\mu = 0$ but
$\tilde s_b \tilde s_{ab} \bar C_\mu = - \partial_\mu \rho \neq 0,
\tilde s_{ab} \tilde s_b C_\mu = + \partial_\mu \lambda \neq 0$.
As a consequence, we find that $\{\tilde s_b, \tilde s_{ab} \}
C_\mu \neq 0$ and $\{\tilde s_b, \tilde s_{ab} \} \bar C_\mu \neq
0$. In the literature, it has been mentioned that the above
transformations are anticommuting modulo the gauge transformations
(i.e. $\{\tilde s_b, \tilde s_{ab} \} C_\mu = - \partial_\mu \rho$
and $\{\tilde s_b, \tilde s_{ab} \} \bar C_\mu = + \partial_\mu
\lambda$) (see, e.g. [2]). However, this unpleasant aspect can be
avoided, as we shall see in the next section.

The gauge-fixing and Faddeev-Popov ghost terms of the Lagrangian
density (1) can be separately written as
\begin{eqnarray}
\tilde s_b \Bigl [ - \bar C^\mu \bigl \{ (\partial^\nu B_{\nu\mu}
-
\partial_\mu \phi)
- \frac{1}{2} B_\mu \bigr \} - \bar \beta \bigl (\partial \cdot C
+ 2 \lambda \bigr ) \Bigr ],
\end{eqnarray}
and
\begin{eqnarray}
\tilde s_{ab} \Bigl [ + C^\mu \bigl \{ (\partial^\nu B_{\nu\mu} -
\partial_\mu \phi)
- \frac{1}{2} B_\mu \bigr \} + \beta \bigl (\partial \cdot \bar C
+ 2 \rho \bigr ) \Bigr ].
\end{eqnarray}
The above expressions provide a simple and straightforward proof
for the nilpotent symmetry invariance of the Lagrangian density
(1) because of (i) the nilpotency of the transformations $\tilde
s_{(a)b}$, and (ii) the invariance of the curvature term (i.e.
$\tilde s_{(a)b} H_{\mu\nu\kappa} = 0$) under $\tilde s_{(a)b}$.
However the above gauge-fixing and Faddeev-Popov ghost terms can
{\it never} be expressed as the BRST ($\tilde s_b$) and anti-BRST
($\tilde s_{ab}$) exact form {\it together} because of the
non-anticommutativity property of the above nilpotent
transformations.

\section{Nilpotent and anticommuting (anti-)BRST symmetry transformations}

Here we will show that the previous drawback can be fixed by
introducing a constrained auxiliary field. It can be seen that the
following off-shell nilpotent (i.e. $s_{(a)b}^2 = 0$) (anti-)BRST
symmetry transformations $s_{(a)b}$
\begin{eqnarray}
&& s_b B_{\mu\nu} = - (\partial_\mu C_\nu - \partial_\nu C_\mu),
\qquad s_b C_\mu = - \partial_\mu \beta, \qquad s_b \bar C_\mu =
- B_\mu, \nonumber\\
&& s_b \phi = \lambda, \quad s_b \bar \beta = - \rho, \quad
s_b \bar B_\mu = - \partial_\mu \lambda, \quad
s_b \bigl [\rho, \lambda, \beta, B_\mu, H_{\mu\nu\kappa} \bigr ] = 0,
\label{BRST1}
\end{eqnarray}
\begin{eqnarray}
&& s_{ab} B_{\mu\nu} = - (\partial_\mu \bar C_\nu - \partial_\nu \bar C_\mu),
\qquad
s_{ab} \bar C_\mu = - \partial_\mu \bar \beta, \qquad s_{ab}  C_\mu
= + \bar B_\mu, \nonumber\\
&& s_{ab} \phi = \rho, \;\; s_{ab} \beta = - \lambda, \;\;
s_{ab} B_\mu = + \partial_\mu \rho, \;\;
s_{ab} \bigl [\rho, \lambda, \bar\beta, \bar B_\mu, H_{\mu\nu\kappa} \bigr ]
= 0, \label{antiBRST1}
\end{eqnarray}
are {\it anticommuting} (i.e. $(s_b + s_{ab})^2 \equiv \{s_b,
s_{ab} \} = 0$) in nature if the whole 4D free Abelian 2-form
gauge theory is defined on a constrained surface parametrized by
the following field equation
\begin{equation}
B_\mu - \bar B_\mu -\partial_\mu \phi = 0\label{CF1}.
\end{equation}
In fact, the anti-commutator $\{ s_b, s_{ab} \} B_{\mu\nu} = 0$ is
valid only if the above equation (9) is satisfied. Moreover, it is
straightforward to check that $\{s_b, s_{ab} \} C_\mu = 0$ and
$\{s_b, s_{ab} \} \bar C_\mu = 0$ which were not true for the
nilpotent symmetry transformations (2) and (3) discussed in the
previous section.

The above equation (9) is the analogue of the Curci-Ferrari
restriction [7] which is defined in the context of proving the
anti-commutativity property (i.e. $\{ s_b s_{ab} \} = 0$) of the
nilpotent (anti-)BRST symmetry transformations $s_{(a)b}$ for the
4D non-Abelian 1-form gauge theory. Furthermore, it can be
checked, by exploiting the equations (7) and (8), that the
condition (9) remains invariant under the anti-commuting
(anti-)BRST symmetry transformations (i.e. $s_{(a)b} [B_\mu - \bar
B_\mu -
\partial_\mu \phi] = 0$). The reason behind the existence of the
constrained field equation (9), for the 4D Abelian 2-form gauge
theory, comes from the superfield approach to BRST formalism [4].

We can express (9) as an equation of motion from a single
Lagrangian density. This can be done if we introduce a Lagrange
multiplier field $L_\mu$  in an appropriate BRST invariant
Lagrangian density in the following manner
\begin{eqnarray}
{\cal L}^{(b)} &=& \frac{1}{12} H^{\mu\nu\kappa} H_{\mu\nu\kappa}
+ B^\mu (\partial^\nu B_{\nu\mu}) + \frac{1}{2} (B \cdot B +  \bar
B \cdot \bar B) \nonumber\\ &+& \partial^\mu \bar\beta
\partial_\mu \beta - \frac{1}{2} \partial^\mu \phi \partial^\mu
\phi + (\partial_\mu \bar C_\nu - \partial_\nu \bar C_\mu)
(\partial^\mu C^\nu) \nonumber\\ &+& (\partial \cdot C - \lambda)
\rho + (\partial \cdot \bar C + \rho) \lambda + L^\mu (B_\mu -
\bar B_\mu-\partial_\mu \phi),\label{LB'}
\end{eqnarray}
where the multiplier field $L_\mu$ transforms under BRST
transformation as: $s_b L_\mu = - \partial_\mu \lambda$. This is
consistent with the equations of motion w.r.t. $B_\mu, L_\mu,$
$\bar B_\mu, \phi$, derived from the above Lagrangian density, as
given below
\begin{eqnarray}
&& \partial^\nu B_{\nu\mu} + B_\mu + L_\mu = 0, \quad \bar B_\mu -
L_\mu = 0, \nonumber\\ && B_\mu - \bar B_\mu -\partial_\mu \phi =
0, \qquad \Box \phi + \partial_\mu L^\mu = 0.
\end{eqnarray}
The above equations, ultimately, imply $\partial \cdot B = 0,
\partial \cdot \bar B = 0, \Box \phi = 0, L_\mu = \bar B_\mu$.
The transformation $s_b L_\mu = - \partial_\mu \lambda$ is
consistent with $L_\mu = \bar B_\mu$ if we compare it with the
BRST transformations (7) under which the Lagrangian density (10)
transforms as: $s_b {\cal L}^{(b)} = - \partial_\mu [(\partial^\mu
C^\nu - \partial^\nu C^\mu) B_\nu + \lambda B^\mu + \rho
\partial^\mu \beta ]$.

It is worthwhile to point out that the gauge-fixing and
Faddeev-Popov ghost terms of the Lagrangian density (10) have been
obtained by exploiting the anti-commuting (anti-)BRST symmetry
transformations (7) and (8) as
\begin{eqnarray}
&& s_b s_{ab} \Bigl [2 \beta \bar \beta + \bar C_\mu C^\mu -
\frac{1}{4} B^{\mu\nu} B_{\mu\nu} \Bigr ] = B^\mu (\partial^\nu
B_{\nu\mu}) + B \cdot \bar B + \partial_\mu \bar\beta
\partial^\mu \beta \nonumber\\ && + (\partial_\mu \bar C_\nu - \partial_\nu \bar
C_\mu) (\partial^\mu C^\nu) + (\partial \cdot C - \lambda) \rho +
(\partial \cdot \bar C + \rho) \lambda.
\end{eqnarray}
We have used the constraint field equation (9) to express
\begin{equation}
B \cdot \bar B = \frac{1}{2} (B \cdot B + \bar B \cdot \bar B) -
\frac{1}{2} \partial_\mu \phi \partial^\mu \phi.
\end{equation}
It should be noted that (12) cannot be obtained in terms of the
nilpotent transformations (2) and (3) which are non-anticommuting
in nature.

Similarly, we can write the anti-BRST invariant Lagrangian density
as:
\begin{eqnarray}
{\cal L}^{(ab)} &=& \frac{1}{12} H^{\mu\nu\kappa} H_{\mu\nu\kappa}
+ \bar B^\mu (\partial^\nu B_{\nu\mu}) + \frac{1}{2} (B \cdot B +
\bar B \cdot \bar B) \nonumber\\ &+& \partial^\mu \bar\beta
\partial_\mu \beta - \frac{1}{2} \partial^\mu \phi \partial^\mu
\phi + (\partial_\mu \bar C_\nu - \partial_\nu \bar C_\mu)
(\partial^\mu C^\nu)  \nonumber\\ &+& (\partial \cdot C - \lambda)
\rho + (\partial \cdot \bar C + \rho) \lambda + L^\mu (B_\mu -
\bar B_\mu - \partial_\mu \phi).\label{LBbar'}
\end{eqnarray}
Note that, only in the second term of the BRST invariant
Lagrangian density (\ref{LB'}), we have changed $B_\mu \to \bar
B_\mu$ which is consistent with (9). The equations of motion,
derived from the ${\cal L}^{(ab)}$, are
\begin{eqnarray}
&& \partial^\nu B_{\nu\mu} + \bar B_\mu - L_\mu = 0, \quad  B_\mu
+ L_\mu = 0, \nonumber\\ && B_\mu - \bar B_\mu -\partial_\mu \phi
= 0, \qquad \Box \phi + \partial_\mu L^\mu = 0.
\end{eqnarray}
We derive, from the above, the equations $\partial \cdot B = 0,
\partial \cdot \bar B = 0, \Box \phi = 0, L_\mu = - B_\mu$. The
anti-BRST symmetry transformation $s_{ab} L_\mu = - \partial_\mu
\rho$ for the Lagrange multiplier field is consistent with $B_\mu
+ L_\mu = 0$ and the transformations (8). Under the latter
nilpotent transformations, the Lagrangian density (14) transforms
as: $s_{ab} {\cal L}^{(ab)} = -
\partial_\mu [(\partial^\mu \bar C^\nu - \partial^\nu \bar C^\mu)
\bar B_\nu - \rho \bar B^\mu + \lambda
\partial^\mu \bar \beta ]$.
The constraint equation (9) emerges, as an equation of motion,
from both the Lagrangian densities (10) as well as (14) which are
equivalent and (anti-)BRST invariant on the constrained surface
defined by the field equation (9).

\section{Gerbes}

The constraint (\ref{CF1}) is intriguing. A similar type of
constraint appears in the non-Abelian 1-form gauge theories when
we implement the requirement of the anticommutativity of the BRST
and anti--BRST transformations. The latter was introduced first by
Curci and Ferrari \cite{CF} and was definitely related to the
non--Abelian structure of the theory (see \cite{BCR} where the
Curci-Ferrari condition was embedded in the appropriate
geometrical context). In the present case, the structure of the
gauge transformations are definitely Abelian. Therefore the
presence of this constraint calls for a definite novel motivation.
We would like to suggest, in this context, that the rationale
behind (\ref{CF1}) is not to be traced back to the non--Abelianity
of the theory but, rather, to an underlying gerbe structure in the
theory represented by the Lagrangian densities (\ref{LB'})
or (\ref{LBbar'}).

Gerbes form a hierarchy of geometrical structures (over
space--time $M$) whose simplest instance is a line bundle, or
0--gerbe (for a mathematical introduction see
\cite{Giraud,Brylinski,Hitchin}, for physical applications see
\cite{gerbe} and references therein). The next more complicated
case, in the above hierarchy, is a 1--gerbe. This is roughly
speaking a `local' line bundle. The latter is the assignment of a
line bundle for each patch of a covering of $M$, for which a
cocycle condition is required for the quadruple intersections
(rather than for triple ones, which characterizes line
bundles).

A 1--gerbe may be characterized by a triple $(B,A,f)$, formed by
the 2-forms $B$, 1-forms $A$ and 0-forms $f$, respectively
\footnote{Henceforth, it will be convenient to use the more
synthetic language of forms, rather than the component fields,
which have been used earlier in the text.}. These are related in
the following way. Given a covering $\{U_i\}$ of $M$, we associate
to each $U_i$ a two--form $B_i$. On a double intersection $U_i
\cap U_j$, we have $B_i-B_j = dA_{ij}$. On the triple
intersections $U_i \cap U_j\cap U_k$, we must have $A_{ij}+
A_{jk}+A_{ki}= d f_{ijk}$. Finally, on the quadruple intersections
$U_i \cap U_j\cap U_k\cap U_l$, the following integral cocycle
condition must be satisfied:
\be
f_{ijl}-f_{ijk}+f_{jkl}-f_{ikl}= 2\;\pi \;n.\label{cocycle} \ee
This integrality condition will not concern us in our Lagrangian
formulation but it has to be imposed as an external condition.

Two triples, represented by $(B,A,f)$ and $(B',A',f')$
respectively, are gauge equivalent if they satisfy the relations
\be
&&B_i'=B_i+ dC_i,\quad\quad {\rm on} \quad U_i\label{gaugeeq1}\\
&&A_{ij}' = A_{ij} + C_i-C_j + d\lambda_{ij} \quad\quad {\rm on}
\quad U_i\cap U_j\label{gaugeeq2}\\ && f_{ijk}' = f_{ijk}' +
\lambda_{ij}+\lambda_{ki}+ \lambda_{jk} \quad\quad {\rm on} \quad
U_i\cap U_j\cap U_k\label{gaugeeq3} \ee for the one--forms $C$ and
the zero--forms $\lambda$.

The pattern for the higher order gerbes is rather clear. For
instance, the 2--gerbes will be characterized by a quadruple
starting from a 3--form and going down to a 0--form field, etc.

We want now to transfer this geometrical information to field
theory. The field content of a 1--gerbe is clear: it is made up of
a two--form field $B$, a one--form gauge field $A$ and a scalar
field $f$ with the gauge transformations
\be
\delta B = d C,\quad\quad \delta C = C +d\lambda, \quad\quad
\delta f = \lambda. \label{gaugetr} \ee

\section{(anti--)BRST for gerbes}

We wish to define the BRST and anti--BRST transformations for the
above theory. The most general field content is given by the triple $(B,A,f)$.
But since $f$ has 0 canonical dimension and since the essential features
are contained in the couple $(B,A)$, we will consider here only the latter.
The inclusion of $f$ is not difficult but yields more cumbersome formulas.
Let us start from a table that contains the order form and ghost number
of all the fields involved:
\vskip 0.5cm
\begin{tabular}{|l|l|l|l|l|l|l|l|l|l|l|l|l|l|l|}
\hline
field & $B$&$A$&$K$&$\bar K$& $C$&$\bar C$& $\beta$ &$\bar \beta$&
$\lambda$&$\bar \lambda$ &$\rho$&$\bar \rho$&$ g$& $\bar g$\\
\hline
form order & 2&1&1&1&1&1&0&0&0&0&0&0&0&0\\
\hline
ghost number & 0&0&0&0&1&-1&2&-2&1&-1&1&-1&0&0\\
\hline
\end{tabular}
\vskip 0.5cm

The appropriate BRST and anti--BRST transformations turn out to be
\begin{eqnarray}
&& s_b\, B = d C, \quad \quad s_b\, C= -d \beta, \nonumber\\
&& s_b\, A= C + d \lambda,\quad\quad s_b\,\lambda = \beta, \nonumber\\
&& s_b \,\bar C = - K, \qquad s_b\,\bar K = d \rho,  \nonumber\\
&& s_b \,\bar\beta = - \bar \rho,  \qquad s_b\,\bar \lambda= g,   \qquad
s_b\, \bar g= \rho, \label{BRST2}
\end{eqnarray}
together with $s_b [\rho, \bar\rho, g, K_\mu, \beta] = 0$, and
\begin{eqnarray}
&& s_{ab}\, B = d\bar C, \qquad s_{ab}\,\bar C = + d\bar \beta, \nonumber\\
&& s_{ab}\, A = \bar C + d \bar \lambda,   \qquad
 s_{ab} \, \bar \lambda  = - \bar \beta, \nonumber\\
&& s_{ab}\, C = + \bar  K,
\qquad s_{ab} \, K = - d \bar \rho , \qquad \nonumber\\
&& s_{ab}\,\beta = + \rho,    \qquad s_{ab}\,\lambda= - \bar g, \qquad
s_{ab}\, g = - \bar \rho, \label{antiBRST2}
\end{eqnarray}
while $s_{ab} [\bar\beta, \bar g, \bar K_\mu, \rho, \bar\rho]= 0$.

It can be easily verified that $(s_b+s_{ab})^2=0$ if the following constraint
is satisfied:
\begin{equation}
\bar K_\mu - \partial_\mu \bar g = K_\mu - \partial_\mu g.\label{CF2}
\end{equation}
This condition is both BRST and anti--BRST invariant.
It is the analogue of the constraint (\ref{CF1}) above and the analogue
of the Curci--Ferrari condition in non--Abelian gauge theories.

It is also evident that, if we disregard the potential $A$, the
transformations (\ref{BRST2},\ref{antiBRST2}) reduce to
(\ref{BRST1},\ref{antiBRST1}). Therefore the latter is but a particular case
of the transformations introduced in this section. Actions with
the symmetry (\ref{BRST2},\ref{antiBRST2}) as well as the implications
with the superfield formalism \cite{BT,BPT,malik3,malik4} will be analyzed
elsewhere.

In the case of a 2--gerbe with field content $(C,B,A,f)$ with
order form (3,2,1,0) respectively (and ghost number zero),
it is not hard to verify that in order to satisfy $(s_b+s_{ab})^2=0$ one
has to impose two constraints
\be
&& H -\bar H + d(K-\bar K)=0, \quad\quad \bar K- K = d(\bar g - g).\label{CF3}
\ee
where $(H,K,g)$ as well as the corresponding barred fields are
(2,1,0)--form field, respectively, with ghost number 0. It is not hard to imagine
how this will generalize to higher order gerbes. This shows,
in particular, that
such constraints as (\ref{CF1},\ref{CF2},\ref{CF3}) are strictly linked
to the gerbe structure.

\section{Discussion}

The condition $(s_b+s_{ab})^2=0$ is a condition that one should
always require. We recall the geometrical interpretation of the BRST transformation
in \cite{BCR}. In non--Abelian gauge theories a BRST transformation is just
an alias for the set of all the gauge transformations. The nilpotency of $s_b$
represents the consistency which is required upon doing two gauge
transformations in different orders. The anti--BRST transformation
represents an independent version of the same operation, therefore it must
be nilpotent too. But considered together, a BRST and an anti--BRST
are just another way to represent the set of gauge transformations.
Therefore they must satisfy collective nilpotency, i.e. $s_b+s_{ab}$ must
be nilpotent, so that, in particular, $s_bs_{ab}+s_{ab} s_b=0$.

This interpretation holds also for the transformations considered in this
paper. It follows that the constraints (\ref{CF1},\ref{CF2},\ref{CF3}) have to
be imposed for consistency. The question that remains to be clarified
is their geometrical meaning, if any. In \cite{BCR} the Curci--Ferrari constraints
for non--Abelian gauge theories were put in the appropriate geometrical
context, but a geometrical interpretation is still lacking. We do not have
a coherent geometrical interpretation of (\ref{CF1},\ref{CF2},\ref{CF3})
either. However we would like to make some remarks.

First, looking at (\ref{CF1}) we notice that it defines a
De Rham cohomology class, represented by the one--form $B_\mu$. Second,
$\phi$ is a nontrivial cocycle of $s_b+s_{ab}$. Third $\phi$ appears
in degree two starting from $B_{\mu\nu}$. Similar things can be said
about (\ref{CF2}), changing $B_\mu$ with $K_\mu$ and $\phi$ with
$g-\bar g$. Therefore (\ref{CF1},\ref{CF2}), and likewise (\ref{CF3}),
look like transgression relations. It would be very interesting
to obtain a complete picture of the geometry behind these relations.

\vskip 1cm

\noindent
{\bf Acknowledgements}\\

\noindent L.B. would like to acknowledge a useful discussion he had with E.Aldrovandi
concerning gerbes. R.P.M. would like to thank SISSA, Trieste,
Italy for the warm hospitality extended to him during his stay at
SISSA. This research was supported for L.B. by the Italian MIUR
under the program ``Superstringhe, Brane e Interazioni
Fondamentali''.

\end{document}